\documentclass[prl,aps,superscriptaddress,showpacs,twocolumn]{revtex4}

\usepackage{bbm}
\usepackage{amsmath}
\usepackage{amssymb}
\usepackage[dvips]{graphicx}
\usepackage{type1cm}
\DeclareGraphicsExtensions{.ps}
\renewcommand{\>}{\rangle}
\newcommand{\<}{\langle}
\DeclareMathOperator{\trace}{tr}

\newcommand{\1}{\mathbbm{1}}

\begin{document}

\title{Sequential Implementation of Global Quantum Operations}

\date{\today}

\author{L. Lamata}
\affiliation{Max-Planck-Institut f\"ur Quantenoptik, Hans-Kopfermann-Strasse 1, 85748 Garching, Germany}
\affiliation{Instituto de F\'{\i}sica Fundamental,
CSIC, Serrano 113-bis, 28006 Madrid, Spain}

\author{J. Le\'{o}n}
\affiliation{Instituto de F\'{\i}sica Fundamental,
CSIC, Serrano 113-bis, 28006 Madrid, Spain}

\author{D. P\'erez-Garc\'{\i}a}
\affiliation{ Dpto. An\'alisis Matem\'atico, Universidad Complutense
de Madrid, 28040 Madrid, Spain}

\author{D. Salgado}
\affiliation{Dpto. Ingenier\'{\i}a Inform\'atica, Universidad Antonio de 
Nebrija, Pirineos 55, 28040 Madrid, Spain}

\author{E. Solano}
\affiliation{Departamento de Qu\'{\i}mica F\'{\i}sica, Universidad del Pa\'{\i}s Vasco - Euskal Herriko Unibertsitatea, Apdo. 644, 48080 Bilbao, Spain}

\begin{abstract}
We study the possibility for a global unitary applied on an arbitrary number of qubits to be decomposed in  a {\it sequential} unitary procedure, where an ancillary system is allowed to interact only once with each qubit. We prove that sequential unitary decompositions are in general impossible for genuine entangling operations, even with an infinite-dimensional ancilla, being the controlled-NOT gate  a paradigmatic example. Nevertheless, we find particular nontrivial operations in quantum information that can be performed in a sequential unitary manner, as is the case of quantum error correction and quantum cloning.
\end{abstract}


\maketitle

Unitary evolution, a fundamental pillar of quantum dynamics, can be described as the action of a time-dependent unitary operator on an initial quantum state. In quantum information, the resulting evolution of a global quantum operation acting on $N$ qubits can be rewritten as a quantum circuit made out of one- and two-qubit gates. The application of these gates is nonsequential in general, and the same qubits can be operated repeatedly at different stages of the prescribed quantum algorithm~\cite{NielsenChuang}. An arbitrary global unitary is able to produce a unitary transformation that maps $N$ qubits to $N$ qubits. In cases where $M$ of the initial $N$ qubits are unknown and the rest are blanks, as happens in quantum cloning~\cite{Scarani} and quantum error correction protocols~\cite{Preskill}, an $M \rightarrow N$ qubit map that is formally described by an isometric operation can be considered. An isometry is a unitary operation acting only on a subset of its domain.

Engineering a realistic arbitrary global unitary $U$ acting simultaneously on $N$ qubits is known to be a difficult problem. It may be  then desirable to decompose $U$ into a sequential unitary procedure, where each qubit is allowed to interact locally and only once with an itinerant ancillary system $A$, and without measurements (see Fig.~1). The picture of a sequential quantum factory where a global U is decomposed into a sequence of local qubit-ancilla unitary steps is a valid one and could be related to particular quantum Turing machines~\cite{Deutsch}. However, one should be careful to differentiate this distinct physical problem with the already studied sequential quantum factory of states~\cite{Schoen1,Schoen2}. The latter was shown to be always possible for {\it any} state, as long as the dimension of the ancilla matches the bond dimension of its corresponding Matrix Product State (MPS) representation~\cite{PerezGarcia}. It is noteworthy to mention the unexpected connection to MPS theory, which has recently shown its power for understanding spin chain correlations~\cite{Fannes}, classical simulation of quantum entangled systems~\cite{Vidal}, and density-matrix renormalization group techniques~\cite{Verstraete}.

\begin{figure}[t!]
\vspace*{0.6cm}
\begin{center}
\includegraphics[width=5cm]{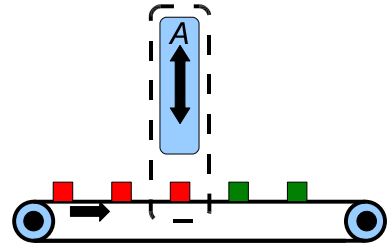}
\end{center}
\vspace{0cm} \caption{Sketch of the proposed sequential quantum factory of operations. When possible, it decomposes a global unitary operation $U$ acting on $N$ qubits into a sequence of local unitary steps where each red qubit interacts only once with the ancilla $A$, and measurements are not allowed. When each qubit-ancilla interaction is finished, the corresponding red qubit in the chain turns into green and the following red qubit starts its interaction with $A$.} \label{stateEvolution}
\end{figure}

In general, sequentiality and the absence of measurements may reduce enormously the complexity in the physical implementation of a global unitary, though they are not unavoidable requirements. However, different experiments and physical settings demand or are restricted to sequential operations for achieving specific quantum information tasks, and allowing measurements may lower the quality of deterministic protocols. Paradigmatic examples are the sequential generation of multi-qubit photonic chains, entangled in energy or polarization states~\cite{Schoen2,Rempe}, and the sequential generation of electron-spin entangled states in quantum dot setups~\cite{Christ}. In the former example, the ancillary system can be a cavity mode, while in the latter it can be a mixed nuclear spin bath. We point out that our proposal is related to the {\it quantum interface} problem~\cite{Rempe}, studying the faithful transfer of information from one quantum node to another. From the point of view of unitary operations, Delgado {\it et al.}~\cite{Delgado} proposed recently a sequential implementation of quantum cloning from $1$ to $N$ copies, but in their scheme qubit measurements were allowed to achieve the desired goal. The case of one-way quantum computing~\cite{Raussendorf} is special, it is deterministic and sequential, but it is measurement based and requires the generation of a particular initial entangled state. Below, when we refer to a sequential decomposition we assume that it is a unitary one, unless other specification is given.

Strikingly, we will prove below that a sequential decomposition of a global unitary $U$ acting simultaneously on $N$ qubits is impossible in general, even with an infinite-dimensional ancilla. By using MPS theory, we will also show how to discriminate the global operations that can be decomposed in a sequential manner and, in the affirmative case, provide with the optimal dimension of the ancilla. We will illustrate our findings with the Shor nine qubit code \cite{Shor} and the optimal $1 \rightarrow N$ cloning machine \cite{Scarani}, showing that the optimal dimension needed for the ancilla in both cases is rather small.

 We first illustrate this surprising result with such a simple example as the paradigmatic two-qubit controlled-NOT (CNOT) gate, which flips the target qubit whenever the control qubit is in state $| 1 \rangle$ and leaves it unchanged when the control qubit is in state  $| 0 \rangle$. We consider a CNOT gate, where qubits $1$ and $2$ are the control and target qubits, respectively, acting on the initial tripartite state $|a \rangle_1 | b \rangle_2 |c \rangle_a$. The latter element in this tripartite product state represents an ancillary system $A$ with arbitrary dimension $D$.

  If we could decompose the CNOT sequentially, we could write
$${\rm CNOT}_{12}\otimes \mathbbm{1}_a=U_{2a} U_{1a}.$$ Therefore, the
input $|+00\>_{12a}$, $|+\>$ being $\frac{1}{\sqrt{2}}(|0\>+|1\>)$, would transform into $\frac{1}{\sqrt{2}}(|00\>+|11\>)_{12}|0\>_a$. This
implies (looking at the entanglement in the partition $1|2a$) that
$U_{1a}|+ 0\>$ has to be maximally entangled. However, $|+
+ 0\>_{12a}$ remains untouched by the CNOT, which implies (again
looking at the partition $1|2a$) that $U_{1a}|+ 0\>$ is a product
state, giving the desired contradiction.

We prove now the following generalization

{\it Theorem 1:} No entangling unitary can be implemented in a sequential way.

Let us call $n$ the smallest $i$ for which $U_{ia}\cdots
U_{1a}=U_{1\ldots i}\otimes U_a$, that is, the smallest step in
which the ancilla always decouples from the qubit chain. This number exists since it
happens at least in the $N$-th step. To finish the argument it is
enough to show that $n=1$. If it were not true, we could group the first
$n-1$ spins and the $n$-th one and then
 have $U_{2a}U_{1a}=U_{12}\otimes \mathbbm{1}_a,$ where $U_{1a}$
entangles $1$ and $a$ (by the choice of $n$). Therefore, assuming
without loss of generality that $U_{1a}|00\>$ is entangled, we may
write
\begin{eqnarray}
0= \<1_a|U_{2a} U_{1a} |0j0\>_{12a} = \sum_{i=0}^r|\phi_i\>_1 \<1_a|U_{2a}|ij\>_{a2} ,
\end{eqnarray}
with $r\ge 1$. Since the $|\phi_i\>_1$ are linearly independent, $\<1_a|U_{2a}|ij\>_{a2}=0$ for all $j$ and $i=0,1$, implying that $\det(U_{2a})=0$ and producing the desired contradiction.

It is noteworthy to mention that, in contrast to the previous
negative result, one can {\it always} implement sequentially a $1\rightarrow N$ isometry as long as the ancilla dimension is large enough, see Fig.~2. The way to see it is straightforward: i) we apply locally the desired global isometry on qubit $1$ and $N-1$ blank ancilla qubits, ii) we implement local swap gates between the ancilla qubits and the $N-1$ blank chain qubits. Clearly, the worst case scenario will require an exponentially large ancilla dimension but this fact does not invalidate the previous general claim \footnote{Note that one could implement any full $N \rightarrow N$ unitary if the following protocol were allowed: i) we map locally the chain qubits onto blank ancilla qubits, ii) we apply locally the desired global unitary on the ancilla qubits, iii) we map {\it back} locally the ancilla qubits onto their corresponding chain qubits. However, it is clear that this recipe breaks the sequential condition.}.

\begin{figure}[t!]
\vspace*{0.4cm}
\begin{center}
\includegraphics[width=5cm]{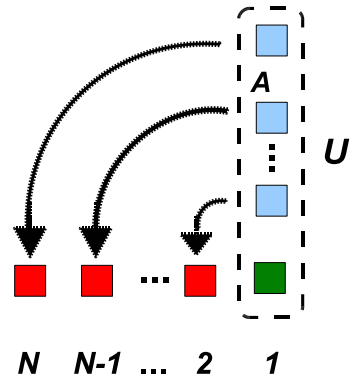}
\end{center}
\vspace{-0.2cm} \caption{Trivial protocol for the sequential decomposition of a $1 \rightarrow N$ isometry. First, the desired isometry is implemented locally between qubit $1$ and $N-1$ qubits in the ancilla degrees of freedom. Second, each ancilla qubit is mapped locally to their corresponding blank chain qubits.} \label{stateEvolution}
\end{figure}

Therefore, the remaining task is to find the optimal dimension of the ancilla needed to implement a given $1 \rightarrow N$ isometry. The solution to this problem will come via the use of MPS~\cite{PerezGarcia} representations of vectors. They can be defined as
\begin{equation}
\label{eq:MPS}
|\psi\>=\sum_{i_1,\ldots,i_N}
A^{[N]}_{i_N}A^{[N-1]}_{i_{N-1}}\cdots
A^{[2]}_{i_{2}}A^{[1]}_{i_1}|i_1 i_2\cdots i_N\rangle,
\end{equation}
where $A^{[m]}_i$ are $D_{m+1}\times D_{m}$ matrices with
$D_1=D_{N+1}=1$. We define the  MPS {\it bond dimension} by
$D=\max_i D_i$ and it can be shown~\cite{Vidal,PerezGarcia} that
there exists a canonical form for any MPS that gives the minimal
bond dimension. This MPS canonical form is defined by the
conditions: (i) $\sum_i
A^{[m]^\dagger}_iA^{[m]}_i=\mathbbm {1}$,
(ii) $\sum_i A^{[m]}_i\Lambda^{[m-1]}
A^{[m]\dagger}_i=\Lambda^{[m]}, $ for all $1\le m\le N$, and (iii)
$\Lambda^{[0]}=\Lambda^{[N]}=1$ with each $\Lambda^{[m]}$
diagonal, positive, full rank and $\trace{\Lambda^{[m]}}=1$. Since
it is obtained by performing successive Schmidt decompositions,
this canonical form is essentially unique. Moreover, using Theorem 3.1.1' of~\cite{Horn2}, one can see that all canonical forms are related by the condition \begin{equation}\label{eq:can} \tilde{A}_i^{[m]}=V_{m} A^{[m]}_iV_{m-1}^\dagger \end{equation} with $V_0=V_N=1$ and $V_m$ unitary such that $[V_m,\Lambda_m]=0$ for all $m$.

In our case we want to get a canonical MPS representation of the $1 \rightarrow N$ isometry $U$. We do it by artificially grouping the input and output of
the first qubit to get
\begin{equation}\label{eq.vidal}
U=\sum_{j_1,i_1,\ldots,i_N} A^{[N]}_{i_N}A^{[N-1]}_{i_{N-1}}\cdots
A^{[2]}_{i_{2}}A^{[1]}_{i_1,j_1}|i_1\>\<j_1|\otimes |i_2\cdots
i_N\rangle .
\end{equation}

 The key observation is to note
that the relations
\begin{align}
\label{eq:seq-dec}
V^{[1]}|0\>_a|j_1\>&=\sum_{i_i,k} \<k|A^{[1]}_{i_1, j_1}
|k\>_a|i_1\>,\\ \nonumber V^{[k]}|r_k\>_a|0\>&=\sum_{i_k,s_k}
\<s_k|A^{[k]}_{i_k}|r_k\>  |s_k\>_a|i_k\> ,
\end{align}
allow us to identify
the existence of a sequential decomposition of $U$, as in Fig.~3, with an MPS decomposition of $U$ as in
(\ref{eq.vidal}) satisfying the additional properties
\begin{equation}\label{eq:conditions}
\sum_{i_1} A^{[1]\dagger}_{i_1,j_1} A^{[1]}_{i_1,j_1'}=
\delta_{j_1,j_1'},\; \; \; \sum_i
A^{[m]^\dagger}_iA^{[m]}_i=\mathbbm {1}, \; m\ge 2.
\end{equation}

\begin{figure}[t!]
\vspace*{0.4cm}
\begin{center}
\includegraphics[width=9cm]{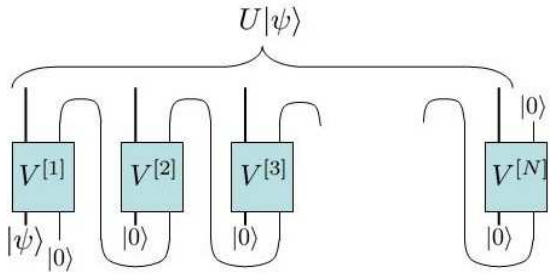}
\end{center}
\vspace{-0.4cm} \caption{Diagram showing the sequential decomposition of a $1\rightarrow N$ isometry $U$, acting on an arbitrary initial state $| \psi \rangle$.} \label{stateEvolution}
\end{figure}

{\it Our main result states that the optimal ancilla to implement
$U$ sequentially is exactly the bond dimension of its {\em canonical} MPS
form. Moreover, any of these canonical MPS representations gives via (\ref{eq:seq-dec}) the sequential decomposition of $U$ as in Fig.~3.}

\

The proof of this result is based on recent mathematical studies in MPS theory~\cite{Vidal,PerezGarcia}.
We start then with a sequential decomposition of our given $1 \rightarrow N$ isometry, for instance the one given by Fig.~2. Using
(\ref{eq:seq-dec}) this gives an MPS representation for $U$, using
matrices $B$. We will show that from there one can obtain a
canonical form for $U$, with matrices $A$, and still satisfying (\ref{eq:conditions}). Using (\ref{eq:can}) this in turn implies that {\it all}  canonical forms for $U$ satisfy (\ref{eq:conditions}).

The procedure to go from matrices $B$ to $A$ was developed in
the proof of Theorem 2 in P\'erez-Garc\'{\i}a {\it et al.}~\cite{PerezGarcia}. There it is shown how to
construct in general non-square matrices $Z_j$ with
$Z_j^{\dagger}Z_j=\mathbbm{1}$ such that, if we define $
A^{[1]}_{i_1,j_1}=Z_1^{\dagger}B^{[1]}_{i_1,j_1}$, $A^{[N]}_{i_N}
=B^{[N]}_{i_N}Z_{N-1} $ and $A^{[m]}_{i_m}
=Z_{m}^{\dagger}B^{[m]}_{i_m}Z_{m-1}$ for $1<m<N$, the $A$'s give
a canonical form for the $U$ operator. Moreover
$Z_1Z_1^{\dagger} B^{[1]}_{i_1,j_1}=B^{[1]}_{i_1,j_1}$. But we finish now by noticing that
\begin{eqnarray}
\sum_{i_1} A^{[1]\dagger}_{i_1,j_1} A^{[1]}_{i_1,j_1'}&=&\sum_{i_1}
B^{[1]\dagger}_{i_1,j_1}Z_1Z_1^{\dagger}
B^{[1]}_{i_1,j_1'}=\delta_{j_1,j_1'} \, , \\
\sum_{i_k}A^{[k]\dagger}_{i_k}A^{[k]}_{i_k}&=&\mathbbm{1}.
\end{eqnarray}

A useful consequence of the previous result is the following

{\bf Corollary:} If $D_i$ is the optimal bond dimension for $U|i\>$, then the optimal ancilla dimension to generate $U$ sequentially is upper bounded by $D_0+D_1$.

\

To see it, let us take MPS
representations of $U|0\>$ and $U|1\>$, given by matrices $A$ and
$B$ respectively. Since $U=U|0\>\<0|+U|1\>\<1|$, to finish the
argument it is enough to notice that one has the MPS decomposition
$U=\sum_{j_1,i_1,\ldots, i_N} C^{[N]}_{i_N}\cdots
C^{[2]}_{i_2}C^{[1]}_{i_1,j_1} |i_1\>\<j_1|\otimes |i_2\cdots
i_N\>,$ taking \footnote{We remark that this
simple construction only gives directly the canonical MPS form
for $U$ in the trivial case $U|0\>=|\psi\>_{1\ldots
N-1}|\varphi\>_N$ and $U|1\>=|\phi\>_{1\ldots
N-1}|\varphi^\perp\>_N$.}
\begin{eqnarray}
C^{[N]}=(A^{[N]}_{i_N}|B^{[N]}_{i_N}),\quad C^{[j]}=\left(
\begin{array}{cc}
  A^{[j]}_{i_j} & 0 \\
  0 & B^{[j]}_{i_j} \\
\end{array}
\right) ,
\end{eqnarray}

\begin{eqnarray}
C^{[1]}_{i_1,0}=\left(
\begin{array}{c}
  A^{[1]}_{i_1} \\
  0 \\
\end{array}
\right),\quad C^{[1]}_{i_1,1}=\left(
\begin{array}{c}
  0 \\
  B^{[1]}_{i_1} \\
  \end{array}
\right) .
\end{eqnarray}

Let us illustrate the strength of the result with a couple of important examples. The first one is the nine-qubit code of Shor \cite{Shor}, which was the first quantum code that allows to correct an arbitrary one-qubit error. It encodes one qubit into nine, via the encoding map $|0\>\mapsto (|000\>+|111\>)^{\otimes 3}$, $|1\>\mapsto
(|000\>-|111\>)^{\otimes 3}$. Since, using the notation of the Corollary, $D_0=D_1=2$, one can implement this
encoding operation sequentially with an ancilla consisting of only
2 qubits (dimension 4). This makes the sequential implementation of
quantum operations appear as an interesting tool for fault
tolerant quantum computation.

The second example is related to optimal quantum cloning~\cite{Scarani}. In this context~\cite{Buzek}, one tries to overcome the restrictions imposed by the no-cloning theorem~\cite{Wootters} at the price of lowering the cloning fidelities. In Ref.~\cite{Gisin}, it is shown how to clone in an optimal way, that is, how to construct a $1\rightarrow (2N-1)$ isometry $U$ such that for any of the first $N$ qubits, the clones, one has (i) $\rho_1=\ldots=\rho_N=\rho$ where $\rho_i$ is the reduced density matrix for qubit $i$ of $U|\psi\>$ (all clones are equal), (ii) the fidelity $\<\psi|\rho|\psi\>$ between the clones and the original state is maximal. For this $U$ it is shown in \cite{Delgado} how $D_0=D_1=N$. Then, using the Corollary, we conclude that the optimal cloning isometry $U$ of one qubit among $N$ participants can be done sequentially with an ancilla of dimension $2N$, that is, consisting of
$\log(N)+1$ qubits.

We have already demonstrated that in the case of a full $N \rightarrow N$ unitary, no entangling global unitary is decomposable in sequential qubit-ancilla local steps, whereas every $1 \rightarrow N$ isometry has
a sequential decomposition. What happens then in the intermediate
case of an $M \rightarrow N$ isometry ($M < N$)? As one could expect, there are
some nontrivial (entangling) unitaries that do admit a sequential
decomposition. To fully characterize which
is the case, one can also apply here the MPS argumentations used in the $1 \rightarrow N$ qubit case to get the following result.

{\it An $M \rightarrow N$ isometry $U$ can be implemented in a sequential
manner if and only if $\sum_{i_k}A^{[k]\dagger}_{i_k,j_k}
A^{[k]}_{i_k,j_k'}=\delta_{j_k,j_k'} \1$ for one (and then for all by Eq. (\ref{eq:can})) canonical form and
for  $1\le k\le M$}.

As a consequence, if $U$ can be implemented in a
sequential manner, the optimal of such decompositions, in terms of the dimension of the ancilla, is given by any of the MPS canonical forms for $U$, written as
\begin{eqnarray} \label{eq:M-N-case}
U=\sum_{j_1,\ldots, j_M,i_1,\ldots,i_N}
A^{[N]}_{i_N}\cdots A^{[M+1]}_{i_{M+1}}A^{[M]}_{i_M,j_M} \nonumber \\ \cdots
A^{[1]}_{i_1,j_1}|i_1\cdots i_N\rangle\langle j_1\cdots j_M|.
\end{eqnarray}

The relations between the sequential decomposition of $U$ as in Fig. 3 and the MPS form (\ref{eq:M-N-case}) are given now by
\begin{align*}
\label{eq:seq-dec}
V^{[1]}|0\>_a|j_1\>&=\sum_{i_i,k} \<k|A^{[1]}_{i_1, j_1}
|k\>_a|i_1\>,\\ V^{[k]}|r_k\>_a|j_k\>&=\sum_{i_k,s_k}
\<s_k|A^{[k]}_{i_k,j_k}|r_k\>  |s_k\>_a|i_k\>\;, \;\;\; k\le M\\
V^{[k]}|r_k\>_a|0\>&=\sum_{i_k,s_k}
\<s_k|A^{[k]}_{i_k}|r_k\>  |s_k\>_a|i_k\>\; , \;\;\; k> M\;.
\end{align*}

In conclusion, we have analyzed the sequential decomposition of arbitrary
unitary operations and we have shown that this is unattainable in
general for the $N \rightarrow N$ qubit case. On the other hand, we
have proved that the $1 \rightarrow N$ qubit case is always
possible and we have found the optimal ancilla dimension. For the intermediate $M \rightarrow N$ qubit case, with
$M < N$, we have given necessary and sufficient conditions for the
sequential decomposition to be possible. We
believe these results are of importance in the context of quantum theory and quantum information applications. With no doubt, they will prove to be relevant in any future implementation of quantum networks, quantum error correction, quantum cryptographic attacks, and related physical setups.

L.L. thanks funding from a Humboldt Research Fellowship. J.L. and L.L. thank support from Spanish projects MEC FIS 2005-05304 and CSIC 2004 5 OE 271, and D. P.-G. from MTM2005-00082 and CCG07-UCM/ESP-2797. D.S. acknowledges support from Spanish Ministry of 
Education project No. 2008-00288 and
E.S. from Ikerbasque Foundation, EU EuroSQIP project, and  UPV-EHU Grant GIU07/40.

\end{document}